\documentclass[preprint,12pt]{elsarticle}
\usepackage{amsmath}
\usepackage{amssymb}
\usepackage{bbm}
\usepackage{amsmath}
\newcommand{\btheta}{\mbox{\boldmath $\theta$}}

\newcommand{\esp}{\end{sloppypar}}
\newcommand{\be}{\begin{equation}}
\newcommand{\ee}{\end{equation}}
\newcommand{\beanno}{\begin{eqnarray*}}

\newcommand{\eeanno}{\end{eqnarray*}}
\newcommand{\bea}{\begin{eqnarray}}
\newcommand{\eea}{\end{eqnarray}}
\newcommand{\ba}{\begin{array}}
\newcommand{\ea}{\end{array}}

\newcommand{\bc}{\begin{center}}
\newcommand{\ec}{\end{center}}

\newcommand{\bpi}{\mbox{\boldmath $\pi$}}
\newcommand{\bSigma}{\mbox{\boldmath $\Sigma$}}

\newcommand{\bbeta}{\mbox{\boldmath $\beta$}}

\newcommand{\bx}{\mbox{\boldmath $x$}}
\newcommand{\bd}{\mbox{\boldmath $d$}}
\newcommand{\bw}{\mbox{\boldmath $w$}}
\newcommand{\bA}{\mbox{\boldmath $A$}}

\usepackage{amssymb}

\journal{Journal of Computational and Applied Mathematics}

\begin{document}

\begin{frontmatter}



\title{Efficient Computational Algorithm for Optimal Allocation in Regression Models}


\author[label1]{Wei Gao}
\author[label2]{Ping Shing Chan}
\author[label3]{Hon Keung Tony Ng}
\author[label1]{Xiaolei Lu}

\address[label1]{Key Laboratory for Applied Statistics of MOE, School of Mathematics and Statistics, Northeast Normal University, Changchun, Jilin 130024, China}
\address[label2]{Department of Statistics, The Chinese University of Hong Kong, Shatin, N. T., Hong Kong}
\address[label3]{Department of Statistical Science, Southern Methodist University, Dallas, Texas 75275, U.S.A. }

\begin{abstract}
In this article, we discuss the optimal allocation
problem in an experiment when a regression model is used for
statistical analysis. Monotonic convergence for a
general class of multiplicative algorithms for $D$-optimality has been discussed in the literature. Here,
we provide an alternate proof of the monotonic convergence for
$D$-criterion with a simple computational algorithm and furthermore
show it converges to the $D$-optimality. We also discuss an algorithm
as well as a conjecture of the monotonic convergence for
$A$-criterion. Monte Carlo simulations are used to demonstrate the
reliability, efficiency and usefulness of the proposed algorithms.
\end{abstract}


\end{frontmatter}



\section{Introduction}
\setcounter{equation}{0}

Regression analysis is an useful technique in modeling and analyzing several variables, when the focus is on the relationship between a dependent variable and one or more independent variables.
It is widely used in different fields of study. For instance, in reliability and life-testing experiments,
often one of the primary purposes is to study the effect of covariates on the failure time distribution and to
 develop inference on the survival probability or some other reliability characteristic of an equipment. For this purpose,
 a regression model is used to incorporate these covariates in the statistical analysis.


Consider a general regression model
\begin{equation}
Y = \mu({\bx}, \bbeta) + \sigma \epsilon \label{RMD}
\end{equation}
where $Y$ is the response variable, $\mu({\bx}, \bbeta)$ is a known function which depends on the unknown parameters $\bbeta \in \Re^{p}$ and the $p$ covariates ${\bx} = ({x}_{1}, \ldots, {x}_{p})^{'}$; and $\epsilon$ is a random variable with $E(\epsilon)= 0$ and $Var(\epsilon) = 1$. We can rewrite the combinations of different levels in different covariates into $k$ experimental conditions (or design points) represent by $\bx_{l} = (x_{1l}~x_{2l}~\ldots~x_{pl})$, where $x_{il}$ is one of the levels of the $i$-th covariate. Note that when the intercept term present in the regression model, $x_{1l} \equiv 1$, $l = 1, \ldots, k$.

Suppose that in an experiment, we have $N$ items available for the test at $k$ experimental conditions. We assign $n_{l}$ items for testing at experimental condition $\bx_{l}$ ($l = 1, 2, \ldots, k$) with $\sum\limits_{l=1}^{k} n_{l} = N$, and observe the corresponding observations for estimation of parameters and/or prediction. In planning such an experiment, we have the flexibility in the choice of ($n_{1}, n_{2}, \ldots, n_{k}$) for a given value of $N$ and the experimental conditions $\bx_{l}, l = 1,2, \ldots, k$. Here, we consider the problem of optimal allocation of $n_1, n_2, \ldots , n_k$ for a general regression model. This optimal allocation problem is usually referred as optimal design problem in the literature.

The optimal design problem in regression setting has long been studied in the literature, for example, Elfving (1952), Fedorov
(1972), Silvey (1980). For extensive developments in optimal design,
one may refer to Silvey (1980), Box and Draper (1987), Atkinson and Donev (1992), Liski et al. (2002), Seber and
Wild (2003) and a
concise introduction by O' Brien and Funk (2003). Besides the rich
development in optimal design theory, different numerical
computational algorithms have been proposed to obtain optimal
designs under different scenarios. For instance, when
$\mu({\bx},\bbeta)$ is linear functions of $\bbeta$, Wynn (1970)
proposed a $W$-algorithm and Fedorov (1972) proposed a $V$-algorithm
to search for the optimal design. Following the ideas in Wynn (1970)
and Fedorov (1972), Mitchell (1974) proposed an algorithm for the
maximization of $|{\bf X}^T{\bf X}|$, where ${\bf X}$ is the design
matrix. Then, Cook and Nachtsheim (1980) provided an empirical
comparison of existing algorithms for the computer generation of
exact $D$-optimal experimental including those due to Wynn (1970),
Fedorov (1972) and Mitchell (1974) and proposed a modification of
the Fedorov (1972) algorithm. However, as pointed out by Silvey
(1980, p.34), these early algorithms have been criticized on the
grounds of their slow convergence and some algorithms have been
suggested to increase the speed of convergence (see, for example,
Atwood, 1973, Silvey and Titterington, 1973 and Wu, 1978). Meyer and
Nachtsheim (1995) proposed a cyclic coordinate-exchange algorithm
for constructing of $D$-optimal designs mainly for continuous design
space. As they stated in Section 2.4, ``For finite design spaces, the
procedure is conceptually simple, although the computational demands
can be prohibitive when $q$ (dimensions of covariates) is large."
Vandenberghe, Boyd and Wu (1998) have proposed the interior-point
method to deal with more general problems but it also suffer from
the slow convergence problem. Recent papers of Torsney and Mandal
(2006), Harman and Pronzato (2007), Dette, Pepelyshev and
Zhigljavsky (2008) and Torsney and Martin-Martin (2009) developed
numerical computational algorithms for $D$-optimal designs. Yu (2010)
discussed the monotonic convergence for a general class of
computational algorithms for $D$-optimal design. In general, it is
desirable to have a numerical computational algorithm to obtain
optimal designs which (i) is simple and reliable; (ii) can be
applied in general situations; and (iii) has a high convergence rate.

In this paper, we aim to develop efficient computational algorithms
to obtain optimal allocation for a general regression model subject
to the $D$-optimality and $A$-optimality criteria. Mathematical results
related to the convergence and monotonicity of the proposed
algorithms are developed. An extensive simulation study is performed
to show the reliability of these algorithms. In Section 2, we
consider the likelihood inference based on a general regression
model and present the forms of the expected Fisher information
matrix and the asymptotic variance-covariance matrix. The two
optimal criteria are also discussed in Section 2. Then, the proposed
computational algorithms for $D$-optimality and $A$-optimality criteria
and their related mathematical properties are discussed in Section
3. Concluding remarks are provided in Section 4. The related proofs of the results are given in Appendix.


\section{Model and Optimal Criteria}

\subsection{Model and Notations}

One of the commonly used approaches to estimate the unknown parameters in a regression model in (\ref{RMD}) is the maximum likelihood method. The maximum likelihood estimates (MLEs) are obtained by maximizing the likelihood function subject to the unknown parameters, $\bbeta$. The properties of the MLEs of the parameters in a regression model are then evaluated based on the asymptotic theory of MLEs. When $\mu({\bx},\bbeta)$ is a linear function of $\bbeta$, the expected Fisher information matrix of the MLE of $\bbeta$ can be expressed as a function of $n_1, \ldots, n_k$,
\begin{equation}
\bSigma(n_1, \ldots, n_k) = \frac{1}{\sigma^{2}} \left[n_1 \bx_{1} \bx_1^{'} + \cdots + n_k \bx_k \bx_k^{'} \right],
\label{SQ}
\end{equation}
where $n_l$ is the number of repeated observations or measurements under the experimental condition $\bx_l$. Thus, the asymptotic variance-covariance matrix of the MLE of $\bbeta$, which is the inverse of the expected Fisher information matrix, can also be expressed as a function of $n_1, \ldots, n_k$.

We can also write the expected Fisher information in terms of $w_{l} = n_{l}/N$, where $w_l = n_l/N$ is the proportion of units (of a total of $N$ units
under test) to be assigned to the experimental condition $\bx_{l}$, $l = 1, 2, \ldots, k$,
\begin{equation}
\bSigma(\bw) = \bSigma(w_1, \ldots, w_k) = N (w_1 \bA_1 + w_2
\bA_2 + \cdots + w_k \bA_k) \label{CQ}
\end{equation}
where $\bA_1, \ldots, \bA_k$ are known nonnegative definite matrices which are functions of $\bx_{1}, \ldots, \bx_{k}$ and $w_l = n_l/N$ is the proportion of units (of a total of $N$ units under test) to be assigned to the experimental condition $\bx_{l}$, $l = 1, 2, \ldots, k$. The related applications that the inverse of covariance is decomposed into sums of linear nonnegative definite matrices has been considered by Vandenberghe, Boyd and Wu (1998), and Qu, Lindsay and Li (2000). The optimal allocation problem is equivalent to obtaining the values of $\bw = (w_{1}, w_{2}, \ldots,
w_{k})$ which optimized a specific objective function subject to the constraints $w_{l} \geq 0$ ($l =1, 2, \ldots, k$) and
$\sum\limits_{l=1}^{k} w_{l} = 1$. It is noteworthy that if the expected Fisher information of the MLEs can be expressed in the form of (\ref{SQ}) or (\ref{CQ}), then the algorithms proposed in this manuscript are applicable. We can show that many of the commonly used regression model, such as
 multiple linear regression model with normal distributed errors, Weibull (extreme-value) regression model and Birnbaum-Saunders regression model,
 which have expected Fisher information of the MLEs in the form of (\ref{SQ}) or (\ref{CQ}). Thus, the proposed algorithms are applicable in those situations.


\subsection{Optimal Criteria}

The goal here is to determine the optimal planning of an experiment when regression analysis is used. We can determine the optimal allocation subject to different optimality criteria. If we are interested in the estimation of the model parameters, we may consider optimality in terms of:
\begin{itemize}
\item[{\bf{[C1]}}] Maximization of the determinant of the Fisher information matrix ${\bSigma}$: This criterion is {\it $D$-optimality}, wherein the determinant of the Fisher information matrix is maximized, which results in minimum volume for the Wald-type joint confidence region for the model parameters $(\bbeta, \sigma)$. $\bw^{*}$ is a $D$-optimal allocation for (\ref{CQ}) if and only if
\begin{equation}
\bw^{*}=\arg\min_{\bw} \left\{-\log(|\Sigma(\bw)|):\;\;\mbox{subject to}\;\;w_l\geq
0\;\;\mbox{and}\;\;\sum\limits_{l=1}^{k}w_l=1 \right\}.\label{DO}
\end{equation}


\item[{\bf{[C2]}}] Minimization of the trace of the variance-covariance matrix (${\bSigma}^{-1}$) of the MLE's: This criterion is {\it $A$-optimality} which minimizes the sum of the variances of the parameter estimates and provides an overall measure of variability from the marginal variabilities. $\bw^{*}$ is a $A$-optimal allocation for (\ref{CQ})
if and only if
{\small
\begin{equation}
\bw^{*}=\arg\min_{\bw}\left\{ \log(\mbox{trace}(\bSigma^{-1}(\bw)):\;\;\mbox{subject
to}\;\;w_l\geq
0\;\;\mbox{and}\;\;\sum\limits_{l=1}^{k} w_l=1\right\}.\label{AO}
\end{equation}}
\end{itemize}

\section{Proposed Computational Algorithm}

In this section, we propose the computational algorithms to obtain the $D$-optimal and $A$-optimal choices of $\bw$. The properties of these algorithms are discussed.

\subsection{Algorithm for $D$-optimal allocation}

\par
\noindent
{\bf Theorem 1.} $\bw^{*}$ is the $D$-optimal choice for (\ref{CQ}) if
and only if
\begin{equation}
\mbox{trace}(\bA_l\bSigma^{-1}(\bw^{*}))=p\;\;\;\mbox{for}\;\;w_l^{*}\not=0\label{D1}
\end{equation}
and
\begin{equation}
\mbox{trace}(\bA_l\bSigma^{-1}(\bw^{*}))\leq
p\;\;\;\mbox{for}\;\;w_l^{*}=0.\label{D2}
\end{equation}
{\bf Proof:} Let $S(\bw)= \log(|{\bf \Sigma}(\bf w)|)$, it is easy
to check that $S(\bw)$ is convex in $\bw$. By Kuhn-Tucker conditions
(Kuhn and Tucker, 1951), ${\bw}^{*}$ is the optimal solution of (4)
if and only if for all $\bw$($w_j\geq 0$ and $\sum w_j = 1$),
 \begin{eqnarray*}
 0 \leq\frac{\partial S({\bw}^{*})} {\partial \bw} (\bw -{\bw}^{*})
 &=& -\sum\limits_{l=1}^{k}\mbox{trace}({\bf
 A}_l{\bf\Sigma}^{-1}({\bw}^{*}))(w_l-w_l^{*})\\
 &=& -\sum\limits_{l=1}^{k}\mbox{trace}({\bf
 A}_l{\bf\Sigma}^{-1}({\bw}^{*}))w_l+p,
\end{eqnarray*}
which implies (\ref{D1}) and (\ref{D2}). Note that Eq. (6) given in Theorem 1 is consistent with the General Equivalence Theorem (Kiefer and Wolfowitz, 1960).

Based on (\ref{D1}) and (\ref{D2}), the following iterative algorithm to obtain the $D$-optimal allocation in (\ref{DO}) is proposed.\\

\noindent {\underline {\bf Algorithm for $D$-optimal allocation}}

\begin{itemize}
\item[{\bf 1.}] Set the initial value of $\bw$ as $\bw^{(0)}=(1/k,\cdots,1/k)^{'}$.
\item[{\bf 2.}] In the $h$-th step, update the value of $\bw$ as
\begin{equation}
w_l^{(h)}=w_l^{(h-1)}\frac{\mbox{trace}(\bA_l\bSigma^{-1}(\bw^{(h-1)}))}{p},
\label{ALGD}
\end{equation}
for $l=1,\cdots,k$. Note that when $\bA_i=\bx_{i}\bx_{i}^{'}$, (\ref{ALGD}) can be expressed as
$$
w_l^{(h)}=w_l^{(h-1)}\frac{\bx_l^{'}\bSigma^{-1}(\bw^{(h-1)})\bx_l}{p}.
$$
\item[{\bf 3.}] Repeat step 2 until the algorithm converge. One of the stopping rule based on absolute difference is stop when $\max \left\{ \left|w_{j}^{(h)} - w_{j}^{(h-1)} \right| \right\} < \zeta$, where $\zeta$ is a small number specified by the user.
\end{itemize}

The monotonic convergence for multiplicative algorithms for $D$-optimality has been established in the literature (see, for example, Yu, 2010). Here, we provide an alternate proof of the monotonic convergence.

\par
\noindent
{\bf Theorem 2.} Let $\{\bw^{(h)}\}$ be given by (\ref{ALGD}), and
then
$$
\log |\bSigma(\bw^{(h)})|-\log |\bSigma(\bw^{(h-1)})|\geq
p\sum\limits_{l=1}^{k}w_l^{(h)}\log\frac{w_l^{(h)}}{w_l^{(h-1)}}\geq\frac{p}{2}\left[\sum\limits_{l=1}^{k}|w_l^{(h)}-w_l^{(h-1)}|\right]^2
$$
and
$$
\bw^{(h)}-\bw^{(h-1)}\rightarrow 0.
$$
\vskip1em
\par
Now we consider the convergence of the proposed algorithm under the condition
\begin{equation}
 a_1A_1+a_2A_2+\cdots+a_kA_k=0 \Longleftrightarrow
 a_1=0,a_2=0,\cdots,a_k=0,
 \label{Co}
\end{equation}
that is, $A_1,\cdots,A_k$ are linearly independent, and this condition is a natural one in order for the models being identifiable.
\par
\noindent
{\bf Theorem 3.} Under the condition of (\ref{Co}), $\{\bw^{(h)}\}$ given by (\ref{ALGD}) is convergent and converges to the
$D$-optimal solution of (\ref{DO}).\\
\par
If $A_1,\cdots,A_k$ are linearly dependent, for the optimal
allocation (\ref{DO}), its solution may not be unique and we can
choose different initial values and get different optimal solutions.
Compared with the W-algorithm proposed by Wynn (1970) and the $V$-algorithm proposed by Fedorov (1972), the value of $\bw$ in the current
step in our proposed algorithm is an explicit function of the value in the previous step which involves simple matrix manipulation while
the value of $\bw$ in each step of the $W$- and $V$-algorithms involve maximizations which required numerical procedures in computation.
Therefore, the algorithms proposed here converge quicker and they are easy to program.

In order to study the convergent rate of the proposed algorithm, an extensive simulation study is performed. We generate the form of the Fisher information matrix in (\ref{CQ}) with the elements of $\bx_1, \bx_2, \cdots, \bx_k$ being independent identically uniform distributed in between $-1$ and 1, i.e., $U(-1, 1)$, for number of design points $k = 10, 20, 30, 40$ and number of covariates $p = 4, 5, 8, 10, 15, 20, 25, 30$ with $p < k$. The stopping criteria of the algorithm are set to be $\max \left\{ |w_{j}^{(h)} - w_{j}^{(h-1)}| \right\} < 0.0001$. For each combination of $p$ and $k$, 50 replications are simulated and their corresponding $D$-optimal allocations are found by using the proposed algorithm. The number of iterations and the elapse time (in unit of second) required to obtain the $D$-optimal allocation are recorded and their average values (with standard deviations in parenthesis) are presented in Table 1.

\begin{table*}[b]
\caption{Simulated results for $D$-optimality} \centering {\small
\begin{tabular}{c c r r c  } \hline
               &      &   &  Average no. of               &  Average elapsed                \\
   $k$         & $p$  &   &  iterations (s.d.)            &  time in sec. (s.d.)            \\
 \hline
10         &     4    &       &        56.8 (27.9)      &   0.192 (0.094) \\
           &     5    &       &        41.9 (16.4)      &   0.143 (0.058) \\
           &     8    &       &        19.5 (10.6)      &   0.067 (0.040) \\
                         \hline
 20        &     4    &       &       96.1 (69.0)       &   0.648 (0.467)  \\
           &     5    &       &       77.9 (29.6)       &   0.534 (0.203)  \\
           &     8    &       &       51.3 (11.5)       &   0.370 (0.085)  \\
           &     10   &       &       36.0 (9.5)        &  0.267 (0.073)   \\
           &     15   &       &       15.8 (4.7)        &  0.129 (0.040)   \\            \hline
 30        &     4     &      &      115.1 (94.8)        &   1.181 (0.982) \\
           &     5     &      &       99.3 (35.9)         &   1.092 (0.539)   \\
           &     8     &      &       60.1 (16.7)         &   0.649 (0.179)    \\
           &     10    &    &         50.2 (14.7)        &   0.565 (0.162) \\
           &     15    &   &          29.0 (4.5)         & 0.354 (0.057)\\
           &     20    &    &         17.9 (3.9)        & 0.255 (0.054)\\
           &    25     & &             9.7 (2.8)         &0.155 (0.046)\\            \hline
  40       &    4     &  &        124.4 (63.9) &        1.673 (0.853)\\
           &    5     &  &       104.7 (38.6) &         1.431 (0.530)\\
           &    8     &  &         72.8 (28.5) &         1.053 (0.419)\\
           &   10     &  &         52.2 (10.2) &         0.791 (0.153)\\
           &   15     &  &         36.7 (6.2)  &       0.608 (0.104)\\
           &   20      &  &         26.0 (6.0)  &     0.497 (0.119)\\
           &   25      &  &         16.7 (3.5)  &     0.362 (0.077)\\
           &   30      &  &        10.7 (2.1)  &     0.273 (0.055)\\
   \hline
\end{tabular}}
\end{table*}

\subsection{Algorithm for $A$-optimal allocation}

\noindent
{\bf Theorem 4.} $\bw^{*}$ is the $A$-optimal choice for (\ref{CQ})
if and only if
\begin{equation}
\frac{\mbox{trace}(\bSigma^{-1}(\bw^{*})\bA_l\bSigma^{-1}({\bw}^{*}))}{\mbox{trace}(\bSigma^{-1}(\bw^{*}))}=1\;\;\;\mbox{for}\;\;w_l^{*}\not=0\label{A1}
\end{equation}
and
\begin{equation}
\frac{\mbox{trace}(\bSigma^{-1}({\bw}^{*})\bA_l\bSigma^{-1}(w^{*}))}{\mbox{trace}(\bSigma^{-1}({\bw}^{*}))}\leq
1\;\;\;\mbox{for}\;\;w_l^{*}=0.\label{A2}
\end{equation}

Based on (\ref{A1}) and (\ref{A2}), the following iterative algorithm to obtain the $A$-optimal allocation in (\ref{AO}) is proposed.\\

\noindent {\underline {\bf Algorithm for $A$-optimal allocation}}

\begin{itemize}
\item[{\bf 1.}] Set the initial value of $\bw$ as $\bw^{(0)}=(1/k,\cdots,1/k)^{'}$.
\item[{\bf 2.}] In the $h$-th step, update the value of $\bw$ as
\begin{equation}
w_l^{(h)}=\frac{w_l^{(h-1)}}{p} \left[\frac{\mbox{trace}(\bSigma^{-1}(\bw^{(h-1)})\bA_l\bSigma^{-1}(\bw^{(h-1)}))}{\mbox{trace}(\bSigma^{-1}(\bw^{(h-1)}))}
+p-1 \right], \label{ALGA}
\end{equation}
for $l=1,\cdots,k$. Note that when $\bA_i=\bx_{i}\bx_{i}^{'}$, (\ref{ALGA}) can be expressed as
$$
w_l^{(h)}=\frac{w_l^{(h-1)}}{p} \left[\frac{\bx_l^{'}\bSigma^{-1}(\bw^{(h-1)})\bSigma^{-1}(\bw^{(h-1)})\bx_l}{\mbox{trace}(\bSigma^{-1}(\bw^{(h-1)}))}
+p-1 \right].
$$
\item[{\bf 3.}] Repeat step 2 until the algorithm converge. One of the stopping rule based on absolute difference is stop when $\max \left\{ |w_{j}^{(h)} - w_{j}^{(h-1)}| \right\} < \zeta$, where $\zeta$ is a small number specified by the user.
\end{itemize}

Although a theoretical justification of convergence of the proposed
computational algorithm for $A$-optimality similar to Theorem 2 is not
yet available, simulation results strongly support the validity and
reliability of the algorithm. An extensive simulation study with
settings presented in Section 3.1 is performed to study the
properties of $A$-optimality. We have generated a wide range of
settings and use our algorithm to compute the $A$-optimal allocation
and we found that the algorithm converge in all these cases. The
number of iterations and the elapse time (in unit of second)
required to obtain the $A$-optimal allocation are recorded and their
average values (with standard deviations in parenthesis) are
presented in Table 2. We conjecture the monotonic convergence of the
algorithm for $A$-optimality and the theoretically prove of
convergence of the proposed computational algorithm for $A$-optimality
seems to be an interesting open problem.
\begin{table*}[b]
\caption{Simulated results for $A$-optimality} \centering {\small
\begin{tabular}{c c r r c  } \hline
               &         &  Average no. of               &  Average elapsed                \\
   $k$         & $p$     &  iterations (s.d.)            &  time in sec. (s.d.)            \\ \hline
10          &     4           &        126.2 (77.0)       &   1.017 (0.618) \\
            &     5           &        121.5 (67.3)       &  0.994 (0.554)  \\
            &     8           &        67.6 (20.7)        &   0.580 (0.180)  \\
                         \hline
 20         &     4           &       188.0 (79.3)         &   3.033 (1.283)  \\
            &     5           &       169.0 (73.9)         &   2.784 (1.218)\\
            &     8           &       119.5 (30.0)         &   2.062 (0.514) \\
            &     10          &       97.5 (18.0)          &  1.739 (0.318)  \\
            &     15          &       72.6 (20.2)          &  1.427 (0.399)     \\
           \hline
 30         &     4           &      218.3 (84.6)           &   5.284 (2.046)   \\
            &     5           &      200.0 (68.0)          & 4.933 (1.682)   \\
            &     8           &       155.8 (38.6)         &4.039 (1.004)    \\
            &     10        &         128.7 (30.0)         &  3.444 (0.809) \\
            &     15       &         92.1 (16.2)          &2.715 (0.475)\\
            &     20        &         76.1 (12.7)         & 2.565 (0.432)\\
            &    25      &           64.5 (12.0)         &2.527 (0.472)\\
            \hline
  40       &    4       &        237.0 (108.6)           &7.635 (3.492)\\
            &    5      &        219.9 (78.0)          &7.416 (2.936)\\
          &    8        &     176.2 (45.6)          &6.078 (1.571)\\
         &   10         &     142.1 (29.3)         &5.072 (1.059)\\
         &   15       &      104.3 (18.8)       &4.114 (0.744)\\
          &   20        &        90.8 (18.8)      & 4.075 (0.842)\\
          &   25        &        77.9 (10.1)      & 4.055 (0.528)\\
          &   30         &        66.4 (8.3)      &4.056 (0.502)\\
   \hline
\end{tabular}}
\end{table*}



\section{Concluding Remarks}

We have proposed simple and efficient iterative algorithms to obtain
the $D$-optimal and $A$-optimal allocations for general regression
model. We have provided an alternate proof of the monotonic
convergence of the proposed algorithm for $D$-optimality and
demonstrate it converges to converges to optimal allocation. We have
also shown that the proposed algorithm for $A$-optimality converges via
an extensive Monte Carlo simulation study and conjecture the the monotonic convergence of
the proposed algorithm for $A$-optimality. The proposed computational
algorithms converges fast and they are easy to program. These
algorithms are programmed in R (R Development Core Team, 2012) and
the programs are available from the authors upon request.

\section*{Appendix: Proof of Theorem 2 and Theorem 3}

\noindent
{\bf Lemma 1.} $\bpi = (\pi_{1}, \pi_{2}, \ldots, \pi_{k})$ and
$\btheta = (\theta_{1}, \theta_{2}, \ldots, \theta_{k})$ are two
probability vectors in $\Re^k$, and then
\begin{equation}
\left[2\sum\limits_{l=1}^{k} \pi_l \log(\pi_l/\theta_l) \right]^{1/2}\geq\sum\limits_{l=1}^{k}|\pi_l - \theta_l|.
\label{INIQ}
\end{equation}
\par \noindent {\bf Proof:} See the proof given by Kullback (1967), Csiszar (1967) or Kemperman (1969). \vskip1em

\par
\noindent
{\bf Lemma 2.} For $d_1\geq 0,\cdots,d_k\geq 0$, let $ \bSigma(\bw,
\bd) = \sum\limits_{l=1}^{k} w_l d_l \bA_l,$ and then
$\bSigma(\bw)=\bSigma(\bw, {\bf 1})$ and
\begin{equation}
\frac{1}{p}\log[|\bSigma(\bw,\bd)|]-\frac{1}{p}\log[|\bSigma(\bw)|]-\sum\limits_{l=1}^{k}\bar{w}_l\log
d_l\geq 0.\label{MIIQ}
\end{equation}
where
$$
 \bar{w}_l=w_l\frac{\mbox{trace}(\bA_l\bSigma^{-1}(\bw))}{p}.
$$
\par
\noindent
{\bf Proof.} Without loss of generality, suppose that $d_1>0,\cdots,
d_k>0$, let $t_1=\log d_1,\cdots,t_k=\log d_k$, and
\begin{eqnarray*}
g(t_1,\ldots,t_k)=\frac{1}{p}\log[|\Sigma(\bw,\exp\{{\bf
t}\})|]-\sum\limits_{l=1}^{k}\bar{w}_lt_l,
\end{eqnarray*}
 and then
$$
\frac{\partial g(t_1,\ldots,t_k)}{\partial
t_l}=w_l\frac{\mbox{trace}(\bA_l\bSigma^{-1}(\bw, \exp\{{\bf
t}\}))}{p}\exp\{t_l\}-\bar{w}_l
$$
and
\begin{eqnarray*}
\frac{\partial^2 g(t_1,\ldots,t_k)}{\partial t_l\partial t_j}&=&-w_l
w_j\frac{\mbox{trace}(\bA_l\bSigma^{-1}(\bw, \exp\{{\bf
t}\})\bA_j\bSigma^{-1}(\bw,\exp\{{\bf t}\}))}{p}\exp\{t_l+t_j\}\\
&&+}w_l\frac{\mbox{trace}(\bA_l\bSigma^{-1}(\bw, \exp\{{\bf
t}\}))}{p}\exp\{t_l\}I_{\{l=j\}.
\end{eqnarray*}

$$
\Gamma=\begin{bmatrix} \displaystyle \frac{\partial^2 g}{\partial
t_1^2}&\displaystyle\frac{\partial^2 g}{\partial t_1\partial
t_2}&\displaystyle\cdots&\displaystyle\frac{\partial^2 g}{\partial t_1\partial t_k}\\ \\ \\
 \displaystyle\frac{\partial^2 g}{\partial t_2\partial t_1}&\displaystyle\frac{\partial^2
g}{\partial t_2^2}&\displaystyle\cdots&\displaystyle\frac{\partial^2
g}{\partial t_2\partial
t_k}\\ \\ \\
 \displaystyle\vdots&\displaystyle\vdots&\displaystyle\ddots&\displaystyle\vdots\\ \\ \\
 \displaystyle\frac{\partial^2 g}{\partial t_k\partial t_1}&\displaystyle\frac{\partial^2
g}{\partial t_k\partial
t_2}&\displaystyle\cdots&\displaystyle\frac{\partial^2 g}{\partial
t_k^2}
 \end{bmatrix}\geq 0
 $$
and ${\partial g(0,\ldots,0)}/{\partial t_l} =0$, and then
$(0,\ldots,0)^{'}$ is the global minimum point and so
$$
g(t_1,\ldots,t_k)\geq g(0,\ldots,0)
$$
which implies that the Lemma holds.

 \vskip1em
\par
\noindent
{\bf Lemma 4.} $\bA$ is a nonnegative matrix and then
\begin{equation}
|\bA|\leq\prod\limits_{i=1}^{p}a_{ii}, \label{DIQ}
\end{equation}
where $a_{ij}$ is the $(i, j)$-th element in $\bA$.
\par
{\bf Proof: } See the results given by Anderson (2003).
 \vskip1em
\noindent
 {\bf Proof of Theorem 2:}
Let $a_{ii}^{(l)}$ be the $i-$th diagonal of ${\bf A}_l$ and by
Lemma 4,
 $$
|\bSigma(\bw^{(h)})|\leq\prod\limits_{i=1}^{p}
\left[\sum\limits_{l=1}^{k}w_l^{(h)}a_{ii}^{(l)}
\right]\leq\prod\limits_{i=1}^{p}\max\{a_{ii}^{(l)};\;\;l=1,\cdots,k\},
 $$
 that is, the sequence $\{\log|\bSigma(\bw^{(h)})|\}$ is uniformly bounded.
 By Lemma 3, we also know the sequence $\{\log|\bSigma(\bw^{(h)})|\}$
 increasing in $n$, and thus it converges.
 \par
 We have
\begin{eqnarray*}
0 & = & \lim_{h\rightarrow\infty} \left[\log| \bSigma(\bw^{(h)})|- \log
|\bSigma(\bw^{(h-1)})| \right] \\
& \geq & \lim_{h\rightarrow\infty}
p\times\sum\limits_{l=1}^{k}w^{(h-1)}_l\log \left[\frac{\mbox{trace}(\bA_l\bSigma^{-1}(\bw^{(h-1)}))}{p} \right]\geq
0
\end{eqnarray*}
which implies
$$
0=\lim_{h \rightarrow \infty}
\sum\limits_{l=1}^{k}w^{(h-1)}_l \log \left[\frac{\mbox{trace}(\bA_l\bSigma^{-1}(\bw^{(h-1)}))}{p} \right]\geq
\sum\limits_{l=1}^{k}|w_l^{(h)}-w_l^{(h-1)}|
$$
and $\bw^{(h)}-\bw^{(h-1)}\rightarrow 0$.
\vskip1em
 \par
Let $W$ be the set of accumulation points of $\{\bw^{(h)}\}$, that
is, for any ${\bw}^{*}\in W$, there exist subsequence
$\{\bw^{(h_s)}\}$ which satisfies
$$
\lim_{s\rightarrow\infty}\bw^{(h_s)}={\bw}^{*}.
$$
\par
For $1\leq l\leq k$ and $1\leq i_1 < i_2 < \cdots < i_l \leq k$, define
\begin{eqnarray*}
& & D(l:\;i_1,\cdots,i_l) \\
& = &\left\{(x_1,\cdots,x_k)^{'}{\Bigg|}\sum\limits_{j=1}^{k}x_j=1,
x_j>0,j\in\{i_1,\cdots,i_l\};\;x_j=0\;\mbox{otherwise}\right\},
\end{eqnarray*}
and
$$
D(l)=\bigcup_{i_1,\cdots,i_l}D(l:\;i_1,\cdots,i_l).
$$
Thus we have
\begin{equation}
W\subset\bigcup_{l=1}^{k}D(l). \label{Nu}
\end{equation}

\par
{\bf Lemma 5.} If $W\bigcap D(l:\;i_1,\cdots,i_l)$ is not an empty
set, \linebreak ${\bw}^{*}\in W\bigcap D(l:\;i_1,\cdots,i_l)$ satisfies
$$
\log|\bSigma({\bw}^{*})|=\max_{\bw\in
D(l:\;i_1,\cdots,i_l)}\log|\bSigma(\bw)|.
$$
\par
{\bf Proof.} By the definition $W$, there exists a subsequence
$\{\bw^{(h_s)}\}$  which is
$$
\lim_{s\rightarrow\infty}\bw^{(h_s)}={\bw}^{*}.
$$
By Theorem 2,
$\displaystyle\lim_{s\rightarrow\infty}\bw^{(h_s)}-\bw^{(h_s-1)}=0$,
have
$$
w_j^{*}=\lim_{s\rightarrow\infty}w_j^{(h_s)}=\lim_{s\rightarrow\infty}w_j^{(h_s-1)}\frac{\mbox{trace}(\bA_j\bSigma^{-1}(\bw^{(h_s-1)}))}{p}
=w_j^{*}\frac{\mbox{trace}(\bA_j\bSigma^{-1}(\bw^{*}))}{p}
$$
which implies
$$
\frac{\mbox{trace}(\bA_j\bSigma^{-1}(\bw^{*}))}{p}=1,
$$
for $j=i_1,\cdots,i_l$, and the conclusion holds by Theorem 1.

\par
\noindent
{\bf Corollary A.} Under (\ref{Co}), $W\bigcap
D(l:\;i_1,\cdots,i_l)$ has no more than one element.
\par
\noindent
{\bf Proof.} The function $log|\bSigma(\bw)|$ is strictly concave
under (\ref{Co}) and so if there have ${\bw}^{*}$, ${\bw}^{**} \in
W\bigcap D(l:\;i_1,\cdots,i_l)$ which satisfy
$$
\log|\bSigma({\bw}^{*})|=\log|\bSigma({\bw}^{**})|=\max_{\bw\in
D(l:\;i_1,\cdots,i_l)}\log|\bSigma(\bw)|,
$$
we have ${\bw}^{*}={\bw}^{**}.$  So $W\bigcap D(l:\;i_1,\cdots,i_l)$
has no more than one element.

\par
\noindent
{\bf Lemma 6.} Let $\{y_m\}$ be a uniformly bounded sequence in
$R^k$. If $y_m-y_{m-1}\rightarrow 0_k$, as $m\rightarrow\infty$, and
the sequence is not convergent, then there are infinitely many
accumulation points of the sequence, where $0_k$ denotes the
$k$-dimensional zero vector.
\par
\noindent
{\bf Proof:} See the Lemma A.1. given by Shi and Jiang (1998).
\vskip1em

\par
\noindent
{\bf Proof of Theorem 3:} Suppose the sequence $\{\bw^{(h)}\}$ does
not converge under (\ref{Co}), and sequence $\{\bw^{(h)}\}$ has
infinitely many accumulation points by Lemma 6. By (\ref{Nu}),
$$
W=W\bigcap\left(\bigcup_{l=1}^{k}D(l)\right)=\bigcup_{l=1}^{k}\bigcup_{i_1,\cdots,i_l}\left(D(l:\;i_1,\cdots,i_l)\cap
 W\right),
$$
and then the number elements $W$ is less than $2^k-1$ which
contradicts. So under (\ref{Co}), the sequence $\{\bw^{(h)}\}$ is
 convergent. Let ${\bw}^{*}=\lim_{h\rightarrow\infty}\bw^{(h)}$, then
$$
w_j^{*}=\lim_{s\rightarrow\infty}w_j^{(h)}=\lim_{s\rightarrow\infty}w_j^{(h-1)}\frac{\mbox{trace}(\bA_j\bSigma^{-1}(\bw^{(h-1)}))}{p}
=w_j^{*}\frac{\mbox{trace}(\bA_j\bSigma^{-1}(\bw^{*}))}{p}
$$
which implies
$$
\frac{\mbox{trace}(\bA_j\bSigma^{-1}(\bw^{*}))}{p}=1,\;\;\;w_j^{*}\not=0
$$
and
$$
\frac{\mbox{trace}(\bA_j\bSigma^{-1}(\bw^{*}))}{p}\leq
1,\;\;\;w_j^{*}=0.
$$
So the sequence $\{\bw^{(h)}\}$ is convergent and converges to the
$D$-optimal solution of (\ref{DO}) by Theorem 1.

\begin{description}
{\centerline{ \large {\underline {References}}}}

\item{} 1. T.W. Anderson, {An Introduction to Multivariate Statistical Analysis,} Wiley, New York, 2003.

\item{} 2. A.C. Atkinson, A.N. Donev, {Optimum Experimental Designs,} Oxford University Press, Oxford, 1992.

\item{} 3. C.L. Atwood, Sequences converging to $D$-optimal designs of experiments, {Annual of Mathematical Statistics} 1 (1973) 342-352.

\item{} 4. G.E.P. Box, N.R. Draper, {Empirical Model Building and Response Surfaces.} John Wiley \& Sons, New York, 1987.

\item{} 5. R.D. Cook, C.J. Nachtsheim, A Comparison of Algorithms for Constructing Exact $D$-Optimal Designs, {Technometrics} 22 (1980) 315-324.

\item{} 6. I. Csiszar, Information-type measures of difference of probability distributions and indirect observations, {Studia Scientiarum Mathematicaeum Hungarica} 2 (1967) 299-318.

\item{} 7. H. Dette, A. Pepelyshev, A. Zhigljavsky, Improving updating rules in multiplicative algorithms for computing $D$-optimal designs, {Computational Statistics and Data Analysis} 53 (2008) 312-320.

\item{} 8. G. Elfving,  Optimum allocation in linear regression theory, {Ann. Math. Statist.} 23 (1952) 255-262.

\item{} 9. V.V. Fedorov, {Theory of optimal Experiments}, Academic Press, New York, 1972.

\item{} 10. R. Harman, L. Pronzato, Improvements on removing nonoptimal support points in $D$-optimum design algorithms, {Statistics and Probability Letter} 77 (2007) 90-94.

\item{} 11. J.H.B. Kemperman, On the optimum rate of transmitting information, in Probability and Information Theory, Springer-Verlag, Berlin (1969) 126-169.

\item{} 12. J. Kiefer, J. Wolfowitz, The equivalence of two extremum problems, Canadian Journal of Mathematics  (1960) 12 363-366.

\item{} 13. H.W. Kuhn, A.W. Tucker, Nonlinear programming, Proceedings of 2nd Berkeley Symposium, University of California Press: Berkeley (1951) 481-492.

\item{} 14. S. Kullback, A lower bound for discrimination information terms of variation, IEEE Transactions on Information Theory  13 (1967) 126-127.

\item{} 15. E.P. Liski, N.K. Mandal, K.R. Shah, B.K. Sinha, Topics in Optimal Design, Springer, New York, 2002.

\item{} 16. R.K. Meyer, C.J. Nachtsheim, The Coordinate-Exchange Algorithm for Constructing Exact Optimal Experimental Designs, Technometrics 37 (1995) 60-69.

\item{} 17. T.J. Mitchell, An Algorithm for the Construction of $D$-Optimal Experimental Designs, Technometrics 16 (1974) 203-210.

\item{} 18. T.E. O'Brien, G.M. Funk, A Gentle Introduction to Optimal Design for Regression Models. American Statistician 57 (2003) 265-267.

\item{} 19. A. Qu, B. Lindsay, B. Li, Improving generalised estimating equations using quadratic inference functions, Biometrika 87 (2000) 823-836.

\item{} 20. R Development Core Team,  R: A language and environment for statistical computing. R Foundation for Statistical Computing, Vienna, Austria. ISBN 3-900051-07-0, URL: http://www.R-project.org, 2012.

\item{} 21. N.-Z. Shi, H. Jiang, Maximum likelihood estimation of isotonic normal mean with unknown variances, J. Multi. Anal. 64 (1998) 183-196.

\item{} 22. S.D. Silvey, {Optimal Design}, Chapman and Hall, New York, 1980.

\item{} 23. S.D. Silvey, D.M. Titterington,  A geometric approach to optimal design theory, {Biometrika} 60 (1973) 21-32.

\item{} 24. G.A.F. Seber, C.J. Wild, Nonlinear regression, John Wiley \& Sons, Hoboken, New Jersey, 2003.

\item{} 25. B. Torsney, S. Mandal, Two classes of multiplicative algorithms for constructing optimizing distributions, {Computational Statistics and Data Analysis}  51 (2006) 1591-1601.

\item{} 26. B. Torsney, R. Martin-Martin, Multiplicative algorithms for computing optimum designs, {Journal of Statistical Planning and Inference}, 139 (2007) 3947-3961.

\item{} 27. L. Vandenberghe, S. Boyd, S.-P. Wu, Determinant Maximization with Linear Matrix Inequality Constraints, {SIAM Journal on Matrix Analysis and Applications} 19 (1998) 499-533.

\item{} 28. C.-F. Wu, Some iterative procedures for generating nonsingular optimal designs, {Commun. Statist.} 14 (1978) 1399-1412.

\item{} 29. H.P. Wynn, The sequential generation of $D$-optimal experimental designs, {Ann. Math. Statist.} 14 (1970) 1655-1664.

\item{} 30. Y. Yu, Monotonic Convergence of a general algorithm for computing optimal designs, Annual of Statistics 38 (2010) 1593-1606.

\end{description}

\end{document}